\documentclass[iop]{emulateapj}
\usepackage{graphicx}

\usepackage{amsmath}
\usepackage{hyperref}

\hypersetup{colorlinks=true, citecolor=red, filecolor=blue,urlcolor=blue,
linkcolor=blue}

\submitted{Received 2011 July 27; accepted 2012 January 25}

\shorttitle{Galaxy scale lenses in the RCS2: Paper I}
\shortauthors{Anguita et al.}

\begin{document}

\title{Galaxy scale lenses in the RCS2: I. First catalog of candidate strong lenses}

\author{T. Anguita}
\affil{Centro de Astro-Ingenier\'ia, Departamento de Astronom\'ia y Astrof\'isica, Pontificia Universidad Cat\'olica de Chile, Casilla 306, Santiago, Chile}
\affil{Max-Planck-Institut f\"ur Astronomie, K\"onigstuhl 17, 69117 Heidelberg, Germany}

\author{L. F. Barrientos}
\affil{Departamento de Astronom\'ia y Astrof\'isica, Pontificia Universidad Cat\'olica de Chile, Casilla 306, Santiago, Chile}

\author{M. D. Gladders}
\affil{Department of Astronomy \& Astrophysics, University of Chicago, 5640 S. Ellis Avenue, Chicago, IL 60637, USA}

\affil{Department of Astronomy and Astrophysics, Kavli Institute for Cosmological Physics, University of Chicago, 5640 S. Ellis Ave., Chicago, IL 60637, USA}

\author{C. Faure}
\affil{Laboratoire d'Astrophysique, Ecole Polytechnique F\'ed\'erale de Lausanne (EPFL), Obervatoire de Sauverny, CH-1290 Versoix, Switzerland.}

\author{H. K. C. Yee}
\affil{Department of Astronomy and Astrophysics, University of Toronto, 50 St George Street, Toronto, Ontario, M5S 3H4, Canada.}

\author{D. G. Gilbank}
\affil{Department of Physics and Astronomy, University of Waterloo, Waterloo, Ontario, N2L 3G1, Canada.}
\affil{Department of Astronomy and Astrophysics, University of Toronto, 50 St George Street, Toronto, Ontario, M5S 3H4, Canada}

\begin{abstract}
We present the first galaxy scale lens catalog from the second Red-Sequence Cluster Survey. The catalog contains 60 lensing system candidates comprised of Luminous Red Galaxy (LRG) lenses at 0.2 $\lesssim$ $z$ $\lesssim$ 0.5 surrounded by blue arcs or apparent multiple images of background sources. The catalog is a valuable complement to previous galaxy galaxy lens catalogs as it samples an intermediate lens redshift range and is composed of bright sources and lenses that allow easy follow-up for detailed analysis. Mass and mass-to-light ratio estimates reveal that the lens galaxies are massive ($\overline{M}$$\approx$5.5$\times$10$^{\rm 11}$ [$M_{\odot}h^{-1}$]) and rich in dark matter ($\overline{M/L}$$\approx$14 [$M_{\odot}/L_{\odot,B} h$]). Even though a slight increasing trend in the mass-to-light ratio is observed from $z$=0.2 to $z$=0.5, current redshift and light profile measurements do not allow stringent constraints on the mass-to-light ratio evolution of LRGs.

\end{abstract}

\keywords{dark matter - galaxies: elliptical and lenticular, cD - galaxies: evolution - gravitational lensing: strong}

\section{Introduction}

Strong gravitational lensing is one of the most powerful tools in observational cosmology. In particular, strong lensing systems in which both lenses and sources are galaxies can provide insights into several current astrophysical and cosmological topics. For example, the configuration of the lensing-induced multiple images and arcs of a background source and their relative fluxes can be used to study the mass profile of lensing galaxies (e.g., \citealt{kochanekdal04,gavazzi07}) and its evolution (e.g., \citealt{faure11,lagattuta10,treu05}), including the measure of galactic substructure (e.g., \citealt{vegetti09,keeton09}). On the other hand, the magnification induced by lensing allows for the study of intrinsically faint and distant sources (e.g., \citealt{claeskens06,anguita09,koester10,bayliss10}). Furthermore, statistical samples of strong lensing systems can be used to test cosmological models (e.g., \citealt{fedeli08, wambsganss08,schwab10}).

\begin{figure*}[!ht]
 \centering
 \includegraphics[width=14cm]{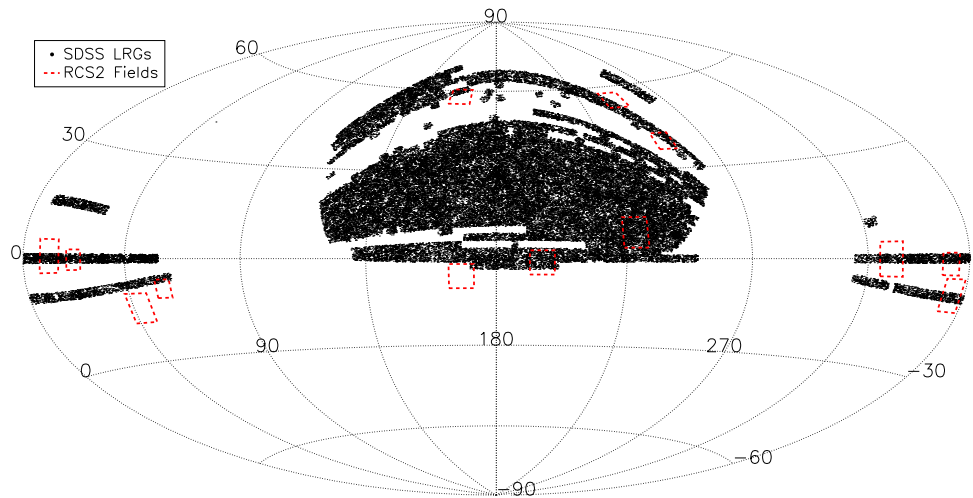}
\caption{RCS2 fields on top of SDSS LRGs. The common area amounts to $\sim$300 deg$^2$ and 2462 cross matches.\label{fig:SDSSRCS}}
\end{figure*}

In order to carry out these studies, a large number of well-sampled strong lensing systems need to be available. Even though galaxy galaxy lensing is a far more common phenomenon than galaxy quasar lensing, since the first confirmation of an extragalactic gravitational lens by \cite{Walsh79}, most of the search efforts have been targeted towards the latter. This is mainly because the emission properties of quasars make multiple imaged quasar systems easier to find and to confirm their lens nature. These efforts have resulted in the discovery of more than 100 multiple imaged gravitationally lensed quasars\footnote{CASTLES Web page (http://www.cfa.harvard.edu/castles/) shows a selection of most of the currently confirmed galaxy quasar lenses.}. Over the last few years, however, with the advent of wide and/or deep imaging and spectroscopic surveys, the number of galaxy galaxy lensing systems discovered have rapidly surpassed the number of galaxy quasar lenses (more than 200 including COSMOS: \citealt{faure08,
jackson08} SLACS: \citealt{bolton08}, GOODS: \citealt{fassnacht04} CFHTLS: \citealt{ruff11} and GEMS: \citealt{more11}). Even when the number counts of lensed galaxies are, and will increasingly be, larger than lensed quasars, galaxy galaxy lensing systems have a few comparative disadvantages regarding individual systems. For example, if searched for in deep space-based imaging surveys (e.g., COSMOS), due to the lower brightness of lensed features, they are considerably harder to follow-up and confirm their nature. On the other hand, when they are searched in spectroscopic surveys (e.g., SLACS), sources are constrained to low redshifts due to the limitations of the spectral range. These ``disadvantages'' however, are, and will be even more so in the future, effectively dealt with by larger sampling, higher data quality and more efficient search methods.

In this paper, we present the first catalog of galaxy scale lenses in the second Red-Sequence Cluster Survey (RCS2). The  RCS2 \citep{gilbank11} is a large optical imaging survey optimized to locate and characterize galaxy clusters over the redshift range 0.1$\lesssim z \lesssim$1.0 using the cluster red-sequence method \citep{gladders00}. The survey comprises 820 deg$^2$ divided into 12 low-extinction patches selected close to the equator (for easy follow-up) and 170 deg$^2$ from the wide component of CFHTLS divided into 4 patches. All these patches were observed with the Canada-France-Hawaii Telescope (CFHT) MegaPrime Camera (MegaCam) in the $g^\prime$, $r^\prime$ and $z^\prime$ filters to a depth of 24.4, 24.3, and 22.8 mag, respectively (5$\sigma$ detections). The MegaCam is composed by a mosaic of 16 CCDs of 4612$\times$2048 pixels each, with a scale of 0''.18 pixel$^{-1}$, yielding a field of view of $\sim$1 deg$^2$.

The paper is organized as follows. We describe our lens search method in Section 2. The properties of the lens catalog are presented in Section 3. In Section 4 we show preliminary results involving lensing mass measurements and mass-to-light ratio evolution and we finish with a summary and discussion in Section 5. For this work, we have assumed a concordance $\Lambda$CDM cosmology with $\Omega_M$=0.3, $\Omega_\Lambda$=0.7, and $H_0$=70 km s$^{-1}$ Mpc$^{-1}$, unless otherwise stated.

\section{The Search Method}

\cite{gilbank11} have created a photometric catalog containing 120 million sources detected in the RCS2 survey. Approximately 87 million of these objects are classified as galaxies, $\sim$8.5 million as stars, and the remaining are either saturated stars, cosmic rays or CCD defects (see \citealt{gilbank11} for details of the survey, data reduction, photometry, and classification). As we are interested in selecting galaxy scale lenses, our search is oriented to arcs or multiple images around objects classified as galaxies in the photometric catalog. However, since it is unrealistic to visually inspect 87 million galaxies, an automated filtering of candidates was done.

Gravitational lensing is a geometric phenomenon whose effectiveness depends on the relative line of sight distances and the projected observed distances (alignment) between the observer, the lens, and the source, as well as the mass of lens. Thus, regarding the properties of lensing galaxies, ideal lens candidates would be very massive early-type galaxies or Luminous Red Galaxies (LRGs). Therefore, we begin by searching for LRGs as potential candidates within the fields of the RCS2 survey given the information available.

\subsection{SDSS-RCS2 common LRGs}

\begin{figure*}[!ht]
 \centering
 \includegraphics[width=16cm]{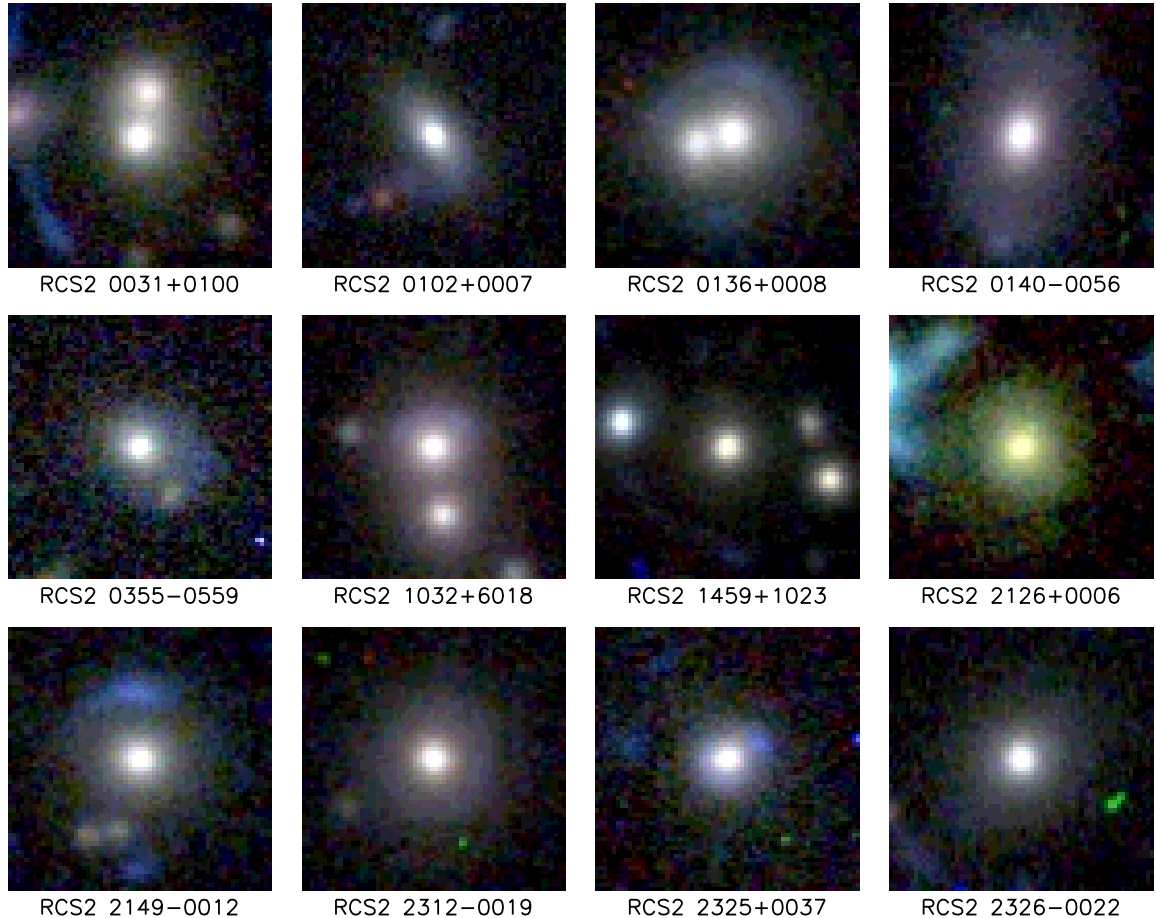}
 \caption{Strong lensing system candidates whose lensing galaxy is a spectroscopically confirmed Luminous Red Galaxy from the SDSS (SDSS LRGs).  The 12 arcsec side length color-composite images were created using the CFHT $g^\prime$,$r^\prime$ and $z^\prime$ RCS2 imaging. \label{fig:SDSSLRGs}}
\end{figure*}

\begin{table*} 
 \renewcommand{\arraystretch}{1.0}
 \centering
  \begin{center}
 \caption{\label{tab:SDSSLRGs} Strong Lensing System Candidates whose Lensing Galaxy is a Spectroscopically Confirmed Luminous Red Galaxy from the SDSS.}
 \begin{tabular}{l r r r r r r r r r}
 \hline
      \multicolumn{1}{c}{\footnotesize Candidate} &  \multicolumn{1}{c}{\footnotesize $z_{\rm l}$}&\multicolumn{1}{c}{\footnotesize $g^\prime$}&\multicolumn{1}{c}{\footnotesize $r^\prime$}&\multicolumn{1}{c}{\footnotesize $z^\prime$}&\multicolumn{1}{c}{\footnotesize $\theta$$_{E}$}&\multicolumn{1}{c}{\footnotesize $R_{E}$}&\multicolumn{1}{c}{\footnotesize Mass}& \multicolumn{1}{c}{\footnotesize $M/L$}& \multicolumn{1}{c}{\footnotesize Grade}\\
     \multicolumn{1}{c}{} & \multicolumn{1}{c}{}&\multicolumn{1}{c}{}&\multicolumn{1}{c}{}&\multicolumn{1}{c}{}&\multicolumn{1}{c}{\footnotesize ('')}&\multicolumn{1}{c}{\footnotesize (kpc)}&\multicolumn{1}{c}{\footnotesize ($\times10^{10}M_{\odot}h^{-1}$)}& \multicolumn{1}{c}{\footnotesize ($M_{\odot}/L_{\odot,B}h$)}& \multicolumn{1}{c}{}\\
\hline
RCS2 0032+0100& 0.390 & 20.92 &    19.21 &    18.22 & 5.8&    30.8&  329.7$^{\rm +  114.7}_{\rm  -43.3}$&  87.0$^{\rm +  30.3}_{\rm -11.4}$&2\\
RCS2 0102+0007& 0.353 & 20.25 &    18.67 &    17.71 & 2.5&    12.4&   54.0$^{\rm +   15.5}_{\rm   -6.3}$&  11.9$^{\rm +   3.4}_{\rm  -1.4}$&1\\
RCS2 0136+0008& 0.344 & 19.43 &    17.86 &    17.02 & 3.3&    15.9&   88.9$^{\rm +   24.5}_{\rm   -9.9}$&  10.0$^{\rm +   2.7}_{\rm  -1.1}$&1\\
RCS2 0140-0056& 0.300 & 19.58 &    18.11 &    17.25 & 4.7&    21.1&  162.4$^{\rm +   35.5}_{\rm  -15.4}$&  34.0$^{\rm +   7.4}_{\rm  -3.2}$&3\\
RCS2 0355-0559& 0.509 & 21.02 &    19.56 &    18.38 & 2.9&    17.9&  114.1$^{\rm +   73.3}_{\rm  -21.3}$&  14.4$^{\rm +   9.3}_{\rm  -2.7}$&2\\
RCS2 1032+6018& 0.400 & 20.14 &    18.51 &    17.54 & 1.6&     8.8&   27.0$^{\rm +    9.9}_{\rm   -3.7}$&   3.4$^{\rm +   1.3}_{\rm  -0.5}$&1\\
RCS2 1459+1023& 0.313 & 20.73 &    19.12 &    18.21 & 4.5&    20.8&  155.4$^{\rm +   36.4}_{\rm  -15.5}$&  73.7$^{\rm +  17.2}_{\rm  -7.4}$&2\\
RCS2 2149-0012& 0.453 & 20.48 &    18.83 &    17.75 & 3.1&    17.9&  111.3$^{\rm +   53.2}_{\rm  -17.7}$&  11.9$^{\rm +   5.7}_{\rm  -1.9}$&3\\
RCS2 2126+0006& 0.506 & 20.66 &    19.16 &    18.00 & 6.0&    37.2&  492.9$^{\rm +  311.1}_{\rm  -91.0}$&  44.2$^{\rm +  27.9}_{\rm  -8.2}$&2\\
RCS2 2312-0019& 0.341 & 20.04 &    18.46 &    17.55 & 2.3&    11.1&   43.2$^{\rm +   11.7}_{\rm   -4.8}$&   8.7$^{\rm +   2.3}_{\rm  -1.0}$&2\\
RCS2 2326-0022& 0.347 & 20.46 &    18.89 &    18.00 & 6.6&    32.3&  367.8$^{\rm +  102.6}_{\rm  -41.7}$& 103.9$^{\rm +  29.0}_{\rm -11.8}$&3\\
RCS2 2325+0037& 0.468 & 20.64 &    19.30 &    18.22 & 1.1&     6.6&   15.3$^{\rm +    7.9}_{\rm   -2.5}$&   2.2$^{\rm +   1.1}_{\rm  -0.4}$&2\\
\hline
\multicolumn{10}{p{.8\textwidth}}{}\\

\multicolumn{10}{p{.8\textwidth}}{\textbf{Notes.} The SDSS spectroscopic redshift of the lens, with a typical uncertainty of $\delta$z=0.001. Shown magnitudes are from the RCS2 photometric catalog; extinction corrected. The mass and mass-to-light ratio estimates are obtained as described in Section 4, assuming a source redshift of $z_s$=1.28 as a central value with a 0.8$<$ $z_s$ $<$2.0 confidence interval. The last column shows the grade assigned to the system on first inspection.}

\end{tabular}

 \end{center}
 \end{table*}

In a first stage, we have cross-matched all the spectroscopically confirmed LRGs from the Data Release 7 of Sloan Digital Sky Survey (SDSS DR7) in RCS2 fields (see Figure \ref{fig:SDSSRCS}), yielding a catalog of 2462 common LRGs with spectroscopic information (hereafter SDSS LRGs). By scaling and subtracting 12 arcsec side length cut-out images in the $g^\prime$ and $z^\prime$ bands of each of the 2462 SDSS LRGs, we eliminated the central galaxy emission while features which do not have the same color of the central galaxy remained. We then searched for bluer multiple images and/or strongly distorted arcs among the remaining features. The cut-out images were selected to be of 12 arcsec side length to limit the mass of our candidate lenses to be of galaxy scale ($<$6 arcsec Einstein radius for a lens at $z_l$=0.3 implies a lens mass within the Einstein radius $\lesssim$ 1$\times$10$^{13}M_\odot$). With a ``fast point and click'' routine, the 2462 systems were identified as either a lens candidate, or 
definitely not a lens, further reducing our sample to $\sim$100 candidate systems. Similarly to what was done by \cite{hennawi08} during this first inspection the $\sim$100 lens candidates were graded according to the following scheme from more likely to less likely:

\begin{itemize}
\item 1: \underline{\textit{Secure lens}}: Typical lens configuration, multiple ($\geq$3) discernible images around the candidate LRG.
\item 2: \underline{\textit{Likely lens}}: Two images or strongly distorted blue arc around the candidate LRG.
\item 3: \underline{\textit{Less likely lens}}: Single arc and/or faint blue feature close to the LRG.
\end{itemize}

We then examined in detail all $\sim$100 candidates using $g^\prime$, $r^\prime$, $z^\prime$ color composite images and further narrowed down the sample to 12 candidate lensing systems (see Table \ref{tab:SDSSLRGs} and Figure \ref{fig:SDSSLRGs}). All of the systems originally graded with 1, most of the systems graded with 2 and a minority of the systems graded with 3 were finally selected.

\subsection{RCS2 LRGs}
\label{sec:methrcs}
\begin{figure}
 \centering
 \includegraphics[clip=true, trim=30pt 0pt 0pt 0pt, width=8.75cm]{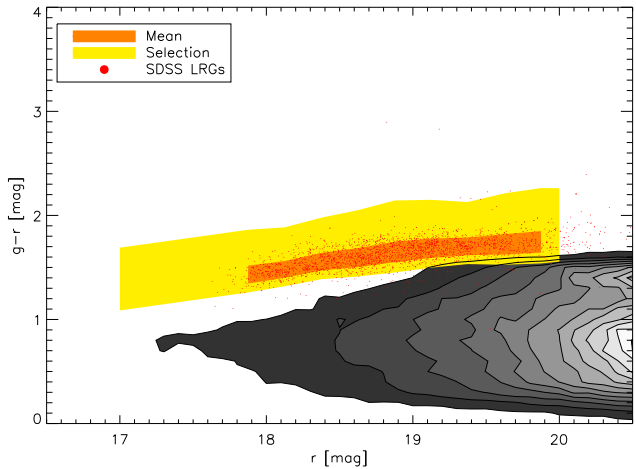}
 \caption{Color-magnitude diagram of the SDSS LRGs with RCS2 photometry (red points). The moving (1$\sigma$) average $g^\prime-r^\prime$ color is shown in orange and the final LRG locus is shown in yellow. The gray scale linear contours show all galaxies in RCS2.}
 \label{fig:CMDsdss}
\end{figure}

The next stage was to search for galaxy scale lenses in all of the RCS2, regardless of common areas with SDSS. As before, to narrow down the search, we selected LRGs as candidate lensing galaxies using the photometric information available. Using the RCS2 catalog, we studied the color-magnitude diagram of the 2462 SDSS LRGs using the RCS2 photometry (see Figure \ref{fig:CMDsdss}). We selected all RCS2 galaxies at the color-magnitude locus defined by

\begin{eqnarray}
 \overline{g^\prime-r^\prime}-2\sigma_{\overline{g^\prime-r^\prime}}<(g^\prime-r^\prime)_{\rm RCS}<\overline{g^\prime-r^\prime}+5\sigma_{\overline{g^\prime-r^\prime}}\\
r^\prime<20,
\label{eq:sellrg}
\end{eqnarray}

\noindent where $\overline{g^\prime-r^\prime}$ is the average $g^\prime-r^\prime$ color per 0.25 mag bin of the SDSS LRGs and $\sigma_{\overline{g-r}}$ is the standard deviation of the average. Even though the fields surveyed by RCS2 are low-extinction patches in the sky, we have corrected the magnitude values for galactic extinction using the maps of \cite{schlegel98} prior to the selection. This first step, yielded a catalog of $\sim$70,000 color-selected LRGs\footnote{It is important to note, that even when throughout this paper these selected galaxies are referred to as ``candidate LRGs'', the catalog should not be considered a complete and thorough LRG sample: the LRGs have only been used as a proxy to search for massive high-redshift lens galaxies (hence the higher cut toward the red part of the color-magnitude diagram and the weak brightness constraint).}.

In a second step, we automatically searched for blue neighbors around each pre-selected LRG. This was done by a simple symmetrization which consisted of subtracting from a $g^\prime$-band stamp image of each galaxy the same stamp image rotated by 180 deg, leaving only the non-symmetric emission within the stamp. The residuals of this subtraction were automatically scanned and those with significant emission ($>$ 4$\sigma$) were selected as more likely lens candidates. As before, the stamp images were selected to be of 12 arcsec side length. This second step reduced the objects in our catalog to $\sim$12,000 candidates which were visually inspected with fast ``point and click'' routines. Approximately 500 systems were found to be possible lenses and graded with the same grading scheme as with the SDSS LRGs. Again, composite color images were created for the $\sim$500 candidates and inspected in detail. This final selection added 48 candidate lensing systems to our catalog (see Table \ref{tab:RCSLRGs} 
and Figure \ref{fig:RCSLRGs1}). We additionally recovered 4 out of the 12 candidate lenses from the SDSS LRGs lens catalog (due to the nature of the automatic part of our search only the brightest non-symmetric systems were recovered).

\subsubsection{Photometric redshifts}

One important difference between the spectroscopically selected LRG candidate lenses and the photometrically selected LRG candidate lenses is that no precise redshift measurements are available for the latter. Even though three-band data would give very unreliable photometric redshifts for a random galaxy, since our candidate lenses are all photometric candidate LRGs, the spectral template used for the redshift estimation can be constrained to be that of an LRG. We have used the SDSS photometric redshift IDL routines \textsc{kphotoz}\footnote{IDL routines can be found at http://howdy.physics.nyu.edu/index.php/Kcorrect} \citep{blanton07} coupled to the extinction-corrected RCS2 $g^\prime$, $r^\prime$ and $z^\prime$ band data of our candidate LRGs, to estimate their redshift. 

To test the accuracy of our photometric redshift measurements, we compare the redshifts obtained using the RCS2 $g^\prime$, $r^\prime$ and $z^\prime$ band data available for the 2462 SDSS LRGs with their measured spectroscopic redshifts (see Fig. \ref{fig:redshifts}). Taking into account a small template shift of $\delta z_{\rm phot}$=0.02, the average dispersion of the photometric redshifts is $\delta z$=0.052. It is important to note the following caveat: this dispersion value is valid for the three-band photometric redshifts of spectroscopically confirmed LRGs. If there are RCS2 galaxies that, even though in the color-magnitude locus of the SDSS LRGs, are not in fact LRGs, they will have an incorrect photometric redshift estimation.

\begin{table*}[!ht] 
 \renewcommand{\arraystretch}{1.0}
 \centering
  \begin{center}
 \caption{\label{tab:RCSLRGs}RCS2 Candidate Photometric LRG Strong Lenses.}
 \begin{tabular}{l r r r r r r r r r}
 \hline
      \multicolumn{1}{c}{\footnotesize Candidate} &  \multicolumn{1}{c}{\footnotesize $z_{\rm l}$}&\multicolumn{1}{c}{\footnotesize $g^\prime$}&\multicolumn{1}{c}{\footnotesize $r^\prime$}&\multicolumn{1}{c}{\footnotesize $z^\prime$}&\multicolumn{1}{c}{\footnotesize $\theta_{E}$}&\multicolumn{1}{c}{\footnotesize $R_{E}$}&\multicolumn{1}{c}{\footnotesize Mass}& \multicolumn{1}{c}{\footnotesize $M/L$}& \multicolumn{1}{c}{\footnotesize Grade}\\
     \multicolumn{1}{c}{} & \multicolumn{1}{c}{}&\multicolumn{1}{c}{}&\multicolumn{1}{c}{}&\multicolumn{1}{c}{}&\multicolumn{1}{c}{\footnotesize ('')}&\multicolumn{1}{c}{\footnotesize (kpc)}&\multicolumn{1}{c}{\footnotesize [$\times10^{10}M_{\odot}h^{-1}$]}& \multicolumn{1}{c}{\footnotesize [$M_{\odot}/L_{\odot,B}h$]}& \multicolumn{1}{c}{}\\
\hline

RCS2 0042+0318	&0.25&18.43&17.17&16.30&   5.1&    19.7& 150.7$^{\rm +    24.6}_{\rm   -11.5}$&  22.2$^{\rm +   3.6}_{\rm  -1.7}$&3 \\ 
RCS2 0036-0027	&0.21&18.48&17.30&16.50&   2.6&     8.9&  33.0$^{\rm +     4.4}_{\rm    -2.1}$&   8.0$^{\rm +   1.1}_{\rm  -0.5}$&2\\
RCS2 0037+0018	&0.29&19.08&17.71&16.83&   3.5&    15.3&  86.1$^{\rm +    17.4}_{\rm    -7.7}$&  14.3$^{\rm +   2.9}_{\rm  -1.3}$&3\\
RCS2 0041+0446	&0.27&19.01&17.65&16.84&   2.3&     9.5&  34.3$^{\rm +     6.2}_{\rm    -2.8}$&   6.5$^{\rm +   1.2}_{\rm  -0.5}$&2\\
RCS2 0101+0313	&0.33&19.66&18.17&17.24&   1.8&     8.5&  25.7$^{\rm +     6.7}_{\rm    -2.7}$&   4.3$^{\rm +   1.1}_{\rm  -0.5}$&2\\
RCS2 0050+0511	&0.40&20.83&19.27&18.20&   2.3&    12.5&  54.1$^{\rm +    19.5}_{\rm    -7.3}$&  14.2$^{\rm +   5.1}_{\rm  -1.9}$&1\\
RCS2 0101-0310	&0.33&20.91&19.36&18.54&   1.2&     5.7&  11.7$^{\rm +     2.9}_{\rm    -1.2}$&   6.1$^{\rm +   1.5}_{\rm  -0.6}$&3\\
RCS2 0034+0243	&0.36&21.33&19.78&18.81&   1.2&     5.9&  12.2$^{\rm +     3.6}_{\rm    -1.5}$&   7.0$^{\rm +   2.1}_{\rm  -0.8}$&3\\
RCS2 0057+0309	&0.37&21.37&19.79&18.80&   1.6&     8.1&  23.0$^{\rm +     7.3}_{\rm    -2.8}$&  12.1$^{\rm +   3.8}_{\rm  -1.5}$&3\\
RCS2 0142+0045	&0.19&18.17&17.05&16.31&   2.5&     7.7&  26.6$^{\rm +     2.9}_{\rm    -1.5}$&   7.3$^{\rm +   0.8}_{\rm  -0.4}$&2\\
RCS2 0126-0225	&0.40&21.03&19.38&18.35&   2.3&    12.4&  53.3$^{\rm +    19.9}_{\rm    -7.4}$&  14.6$^{\rm +   5.4}_{\rm  -2.0}$&3\\
RCS2 0256-1257	&0.37&20.25&18.67&17.68&   1.5&     7.6&  20.2$^{\rm +     6.4}_{\rm    -2.5}$&   3.8$^{\rm +   1.2}_{\rm  -0.5}$&2\\
RCS2 0252-1328	&0.32&20.32&18.87&17.93&   2.5&    11.8&  49.7$^{\rm +    12.2}_{\rm    -5.1}$&  17.2$^{\rm +   4.2}_{\rm  -1.8}$&3\\
RCS2 0255-1451	&0.36&20.54&19.00&18.02&   2.6&    13.3&  61.7$^{\rm +    18.5}_{\rm    -7.4}$&  17.2$^{\rm +   5.2}_{\rm  -2.0}$&2\\
RCS2 0350-0710	&0.31&20.35&18.88&18.03&   2.1&     9.4&  32.2$^{\rm +     7.3}_{\rm    -3.1}$&  12.9$^{\rm +   2.9}_{\rm  -1.3}$&3\\
RCS2 0346-0854	&0.36&20.91&19.27&18.41&   2.9&    14.4&  72.8$^{\rm +    21.2}_{\rm    -8.5}$&  27.3$^{\rm +   8.0}_{\rm  -3.2}$&2\\
RCS2 1032+6018	&0.38&20.14&18.51&17.54&   1.6&     8.3&  24.2$^{\rm +     8.0}_{\rm    -3.1}$&   3.6$^{\rm +   1.2}_{\rm  -0.5}$&1\\
RCS2 1024+5823	&0.33&20.10&18.61&17.69&   1.6&     7.6&  20.6$^{\rm +     5.3}_{\rm    -2.2}$&   5.2$^{\rm +   1.3}_{\rm  -0.6}$&1\\
RCS2 1018+5959	&0.31&20.15&18.71&17.80&   3.4&    15.7&  89.1$^{\rm +    20.7}_{\rm    -8.9}$&  29.2$^{\rm +   6.8}_{\rm  -2.9}$&2\\
RCS2 1042+5831	&0.34&20.23&18.69&17.79&   2.8&    13.6&  65.6$^{\rm +    17.6}_{\rm    -7.2}$&  16.6$^{\rm +   4.4}_{\rm  -1.8}$&3\\
RCS2 1024+5829	&0.34&20.43&18.90&17.98&   1.8&     8.8&  27.1$^{\rm +     7.4}_{\rm    -3.0}$&   8.1$^{\rm +   2.2}_{\rm  -0.9}$&2\\
RCS2 1101+5725	&0.43&21.24&19.58&18.46&   3.0&    16.8&  97.9$^{\rm +    42.4}_{\rm   -14.7}$&  24.8$^{\rm +  10.8}_{\rm  -3.7}$&3\\
RCS2 1109-0912	&0.26&19.19&17.86&17.04&   2.9&    11.9&  53.8$^{\rm +     9.4}_{\rm    -4.3}$&  13.2$^{\rm +   2.3}_{\rm  -1.1}$&3\\
RCS2 1057-0816	&0.33&19.55&18.05&17.16&   2.2&    10.2&  36.8$^{\rm +     9.2}_{\rm    -3.9}$&   5.8$^{\rm +   1.4}_{\rm  -0.6}$&1\\
RCS2 1103-0526	&0.37&21.36&19.81&18.80&   1.6&     8.1&  22.8$^{\rm +     7.2}_{\rm    -2.8}$&  12.4$^{\rm +   3.9}_{\rm  -1.5}$&2\\
RCS2 1101-0602	&0.45&21.44&19.86&18.67&   2.6&    15.1&  79.2$^{\rm +    37.5}_{\rm   -12.5}$&  22.1$^{\rm +  10.5}_{\rm  -3.5}$&1\\
RCS2 1250+0220	&0.42&20.71&19.06&17.98&   1.3&     7.3&  18.6$^{\rm +     7.4}_{\rm    -2.7}$&   3.3$^{\rm +   1.3}_{\rm  -0.5}$&3\\
RCS2 1252+0251	&0.36&20.61&19.07&18.09&   2.0&    10.0&  35.0$^{\rm +    10.4}_{\rm    -4.2}$&  10.5$^{\rm +   3.1}_{\rm  -1.2}$&2\\
RCS2 1511+0642	&0.07&16.19&15.38&14.70&   0.7&     1.0&   0.9$^{\rm +     0.0}_{\rm    -0.0}$&   0.5$^{\rm +   0.1}_{\rm  -0.1}$&2\\
RCS2 1516+1118	&0.32&19.85&18.36&17.50&   2.6&    11.9&  51.1$^{\rm +    12.2}_{\rm    -5.2}$&  11.6$^{\rm +   2.8}_{\rm  -1.2}$&1\\
RCS2 1501+0957	&0.41&20.92&19.29&18.23&   5.8&    31.4& 341.7$^{\rm +   131.6}_{\rm   -47.8}$&  82.0$^{\rm +  31.6}_{\rm -11.5}$&2\\
RCS2 1459+1049	&0.36&21.40&19.82&18.89&   1.6&     8.0&  22.3$^{\rm +     6.6}_{\rm    -2.6}$&  13.8$^{\rm +   4.0}_{\rm  -1.6}$&2\\
RCS2 1522+0640	&0.46&21.33&19.83&18.59&   2.4&    13.9&  67.1$^{\rm +    33.1}_{\rm   -10.9}$&  17.0$^{\rm +   8.4}_{\rm  -2.8}$&2\\
RCS2 2158+0257	&0.31&19.10&17.63&16.78&   2.9&    13.3&  63.9$^{\rm +    14.6}_{\rm    -6.3}$&   8.0$^{\rm +   1.8}_{\rm  -0.8}$&1\\
RCS2 2142+0207	&0.29&20.31&18.85&18.05&   2.1&     9.0&  29.8$^{\rm +     6.2}_{\rm    -2.7}$&  13.3$^{\rm +   2.8}_{\rm  -1.2}$&1\\
RCS2 2136-0056	&0.26&20.27&18.91&18.12&   2.5&    10.2&  39.4$^{\rm +     7.0}_{\rm    -3.2}$&  25.3$^{\rm +   4.4}_{\rm  -2.0}$&2\\
RCS2 2125-0054	&0.30&20.69&19.31&18.35&   2.5&    11.3&  46.3$^{\rm +    10.3}_{\rm    -4.5}$&  28.2$^{\rm +   6.3}_{\rm  -2.7}$&2\\
RCS2 2138-0218	&0.35&21.34&19.79&18.86&   2.6&    12.7&  56.8$^{\rm +    15.8}_{\rm    -6.4}$&  36.9$^{\rm +  10.3}_{\rm  -4.2}$&1\\
RCS2 2313-0225	&0.27&19.84&18.48&17.66&   2.1&     8.5&  27.5$^{\rm +     5.0}_{\rm    -2.3}$&  11.0$^{\rm +   2.0}_{\rm  -0.9}$&1\\
RCS2 2328-0312	&0.25&21.05&19.34&19.02&   3.2&    12.7&  62.4$^{\rm +    10.3}_{\rm    -4.8}$&  66.4$^{\rm +  11.0}_{\rm  -5.1}$&2\\
RCS2 2332-0210	&0.37&21.01&19.51&18.45&   1.5&     7.8&  21.2$^{\rm +     6.7}_{\rm    -2.6}$&   8.7$^{\rm +   2.7}_{\rm  -1.1}$&3\\
RCS2 2333-0355	&0.33&21.15&19.63&18.73&   1.3&     6.4&  14.6$^{\rm +     3.8}_{\rm    -1.6}$&   9.1$^{\rm +   2.4}_{\rm  -1.0}$&2\\
RCS2 2327-1029	&0.33&20.42&18.94&18.02&   1.6&     7.7&  20.9$^{\rm +     5.2}_{\rm    -2.2}$&   7.4$^{\rm +   1.9}_{\rm  -0.8}$&1\\
RCS2 2337-0655	&0.42&20.85&19.17&18.09&   3.8&    21.0& 152.7$^{\rm +    63.1}_{\rm   -22.3}$&  28.7$^{\rm +  11.9}_{\rm  -4.2}$&2\\
RCS2 2339-0859	&0.39&21.17&19.57&18.54&   1.9&    10.2&  36.1$^{\rm +    12.7}_{\rm    -4.8}$&  13.1$^{\rm +   4.6}_{\rm  -1.7}$&3\\
RCS2 2346-0925	&0.38&21.39&19.70&18.79&   2.3&    11.7&  47.7$^{\rm +    15.7}_{\rm    -6.0}$&  21.7$^{\rm +   7.1}_{\rm  -2.7}$&3\\
RCS2 2351-1012	&0.33&21.22&19.70&18.81&   2.0&     9.8&  34.0$^{\rm +     8.8}_{\rm    -3.6}$&  23.0$^{\rm +   6.0}_{\rm  -2.5}$&3\\
RCS2 2348-0613	&0.44&21.62&19.94&18.80&   4.1&    23.5& 191.7$^{\rm +    86.6}_{\rm   -29.5}$&  62.8$^{\rm +  28.4}_{\rm  -9.7}$&2\\
\multicolumn{10}{p{.8\textwidth}}{}\\

\multicolumn{10}{p{.8\textwidth}}{\textbf{Notes.} The redshift column shows the photometric redshift obtained as described in the text. The mean error on the photometric redshift estimates is $\delta z$=0.052. Shown magnitudes are from the RCS2 photometric catalog; extinction corrected. The mass and mass-to-light ratio estimates are obtained as described in the text, assuming a source redshift of $z_s$=1.28 as a central value with a 0.8$<$$z_s$$<$2.0 confidence interval. The last column shows the grade assigned to the system on first inspection.}

\end{tabular}
 \end{center}
 \end{table*}

\begin{figure*}[ht!]
 \centering
 \includegraphics[width=16cm]{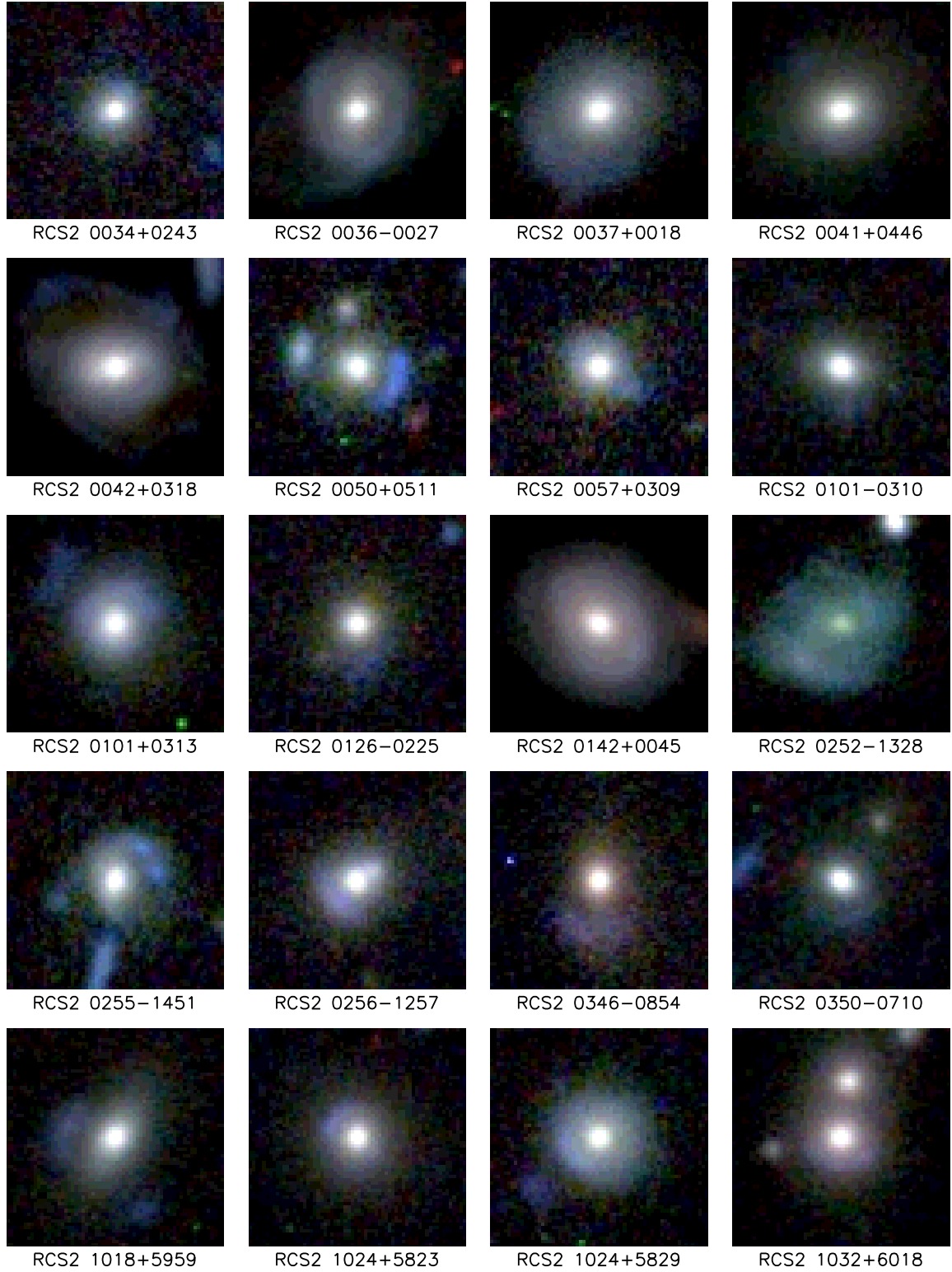}
 \caption{Lensing system candidates whose lensing galaxy is a photometrically selected Luminous Red Galaxy from the RCS2. The 12 arcsec side length color-composite images are created using the CFHT $g^\prime$, $r^\prime$ and $z^\prime$ RCS2 imaging.}
 \label{fig:RCSLRGs1}
\end{figure*}
\addtocounter{figure}{-1}

\begin{figure*}[ht!]
 \centering
 \includegraphics[width=16cm]{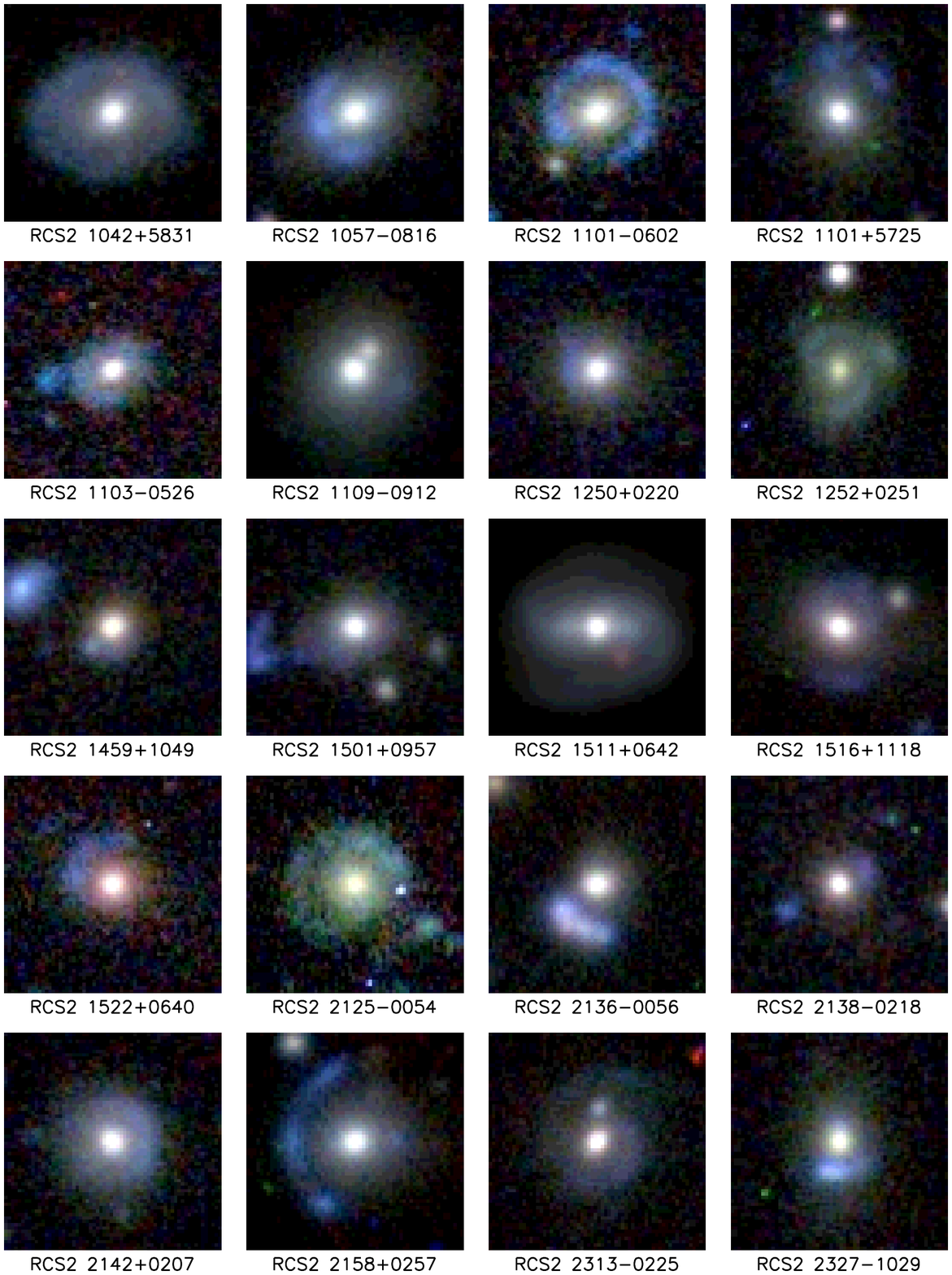}
 \caption{\textit{Continued}}
 \label{fig:RCSLRGs2}
\end{figure*}
\addtocounter{figure}{-1}

\begin{figure*}[ht!]
 \centering
 \includegraphics[width=16cm]{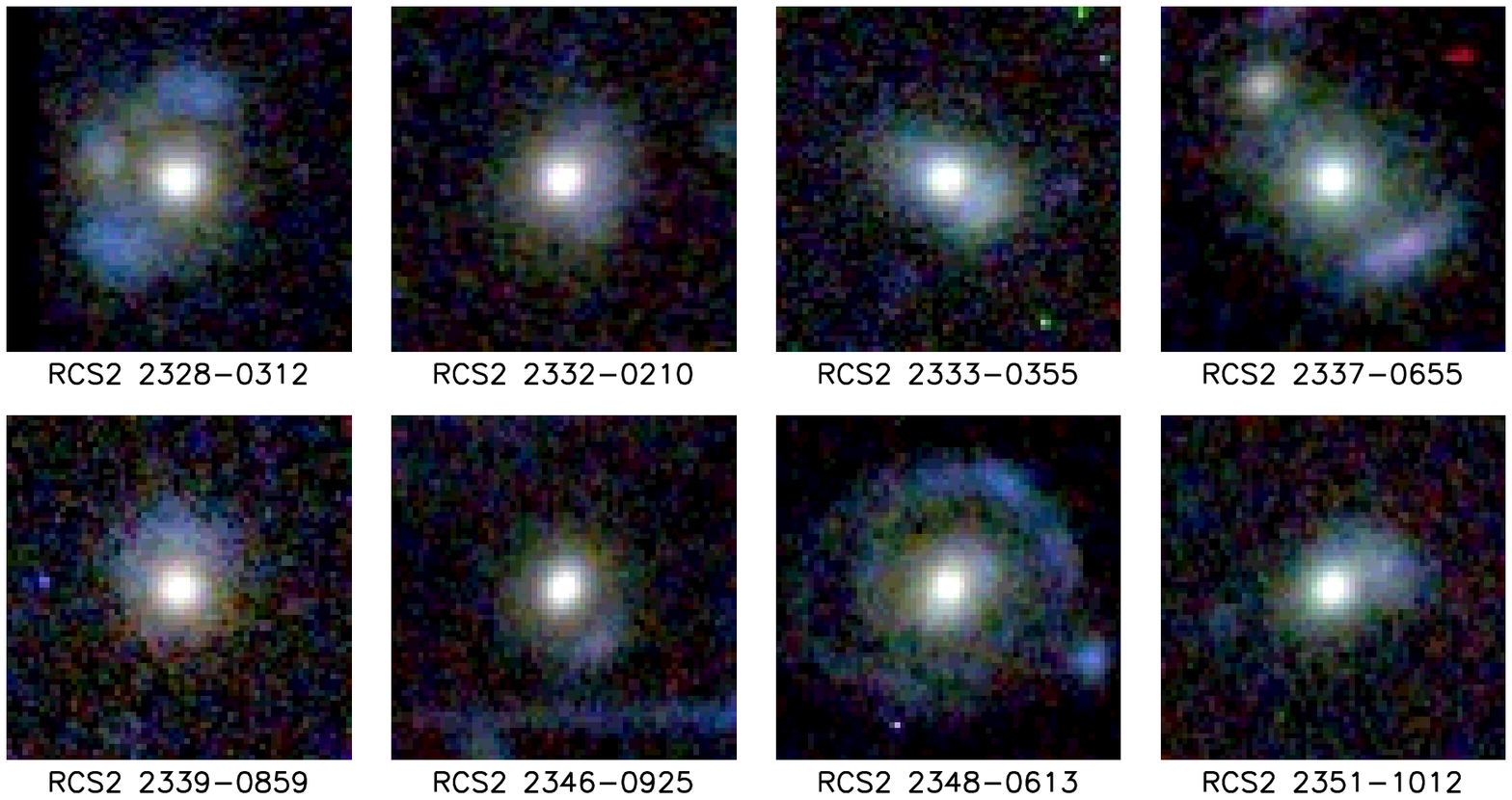}
 \caption{\textit{Continued}}
 \label{fig:RCSLRGs2}
\end{figure*}

\renewcommand{\thefigure}{\arabic{figure}}

\section{The Catalog}                                                                                                   
                                                                                                                        
The complete catalog contains a total number of 60 strong lensing system candidates over a redshift range 0.2$\lesssim$z$\lesssim$0.5, 12 systems with spectroscopically confirmed LRG lenses and 48 with photometric LRG candidate lenses. Their RCS2 photometry, lens redshifts, and coordinates are shown in Tables \ref{tab:SDSSLRGs} and \ref{tab:RCSLRGs}, while composite $g^\prime$, $r^\prime$ and $z^\prime$-band color images are displayed in Figures \ref{fig:SDSSLRGs} and \ref{fig:RCSLRGs1}.

\begin{figure}
 \centering
 \includegraphics[clip=true, trim=30pt 0pt 0pt 0pt, width=8.75cm]{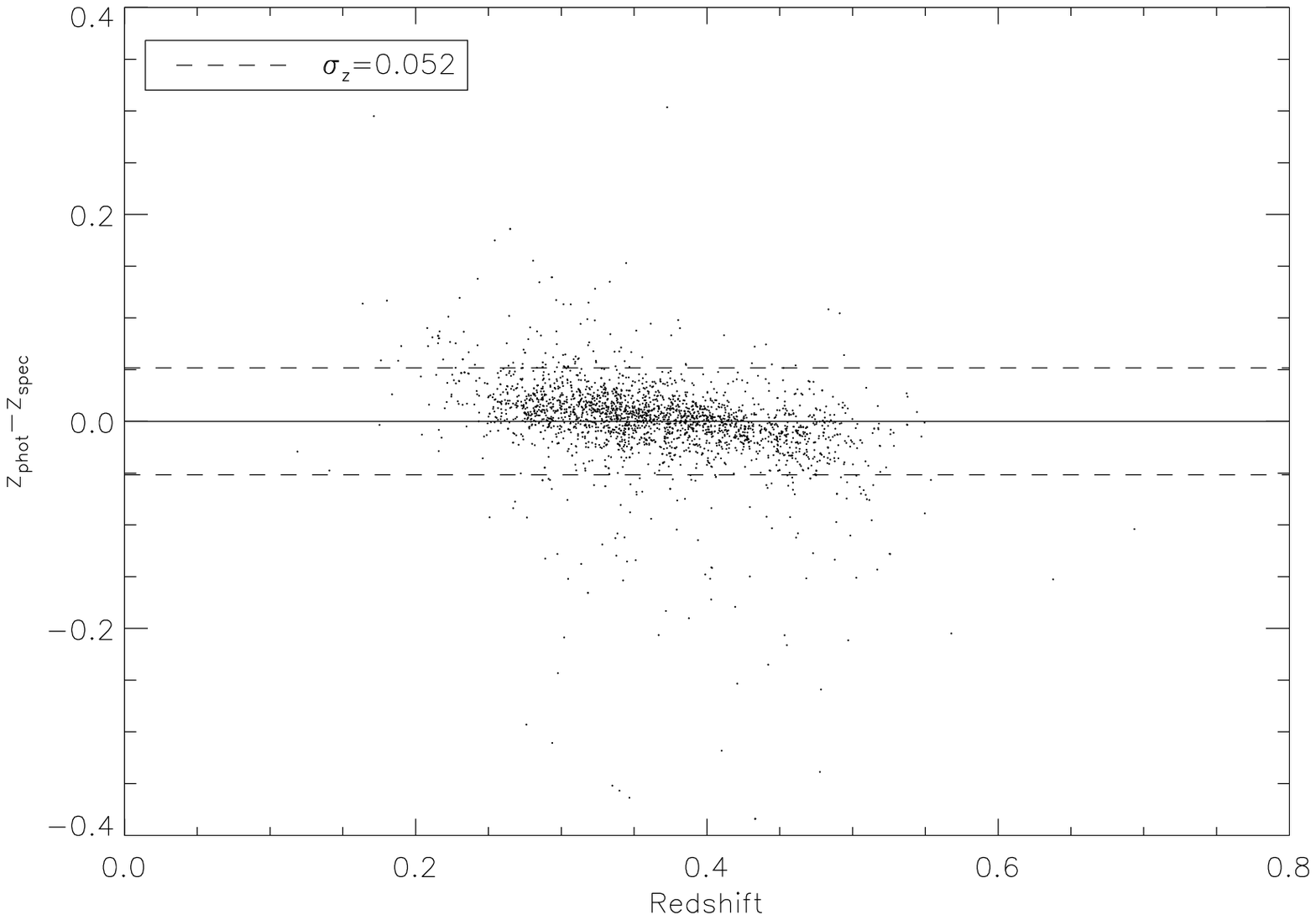}
 \caption{Three-band (RCS2 $g^\prime$, $r^\prime$ and $z^\prime$) photometric redshift accuracy relative to the spectroscopic redshift for the SDSS LRGs.}
 \label{fig:redshifts}
\end{figure}

\begin{figure}[h]
 \centering
 \includegraphics[clip=true, trim=30pt 0pt 0pt 0pt, width=8.75cm]{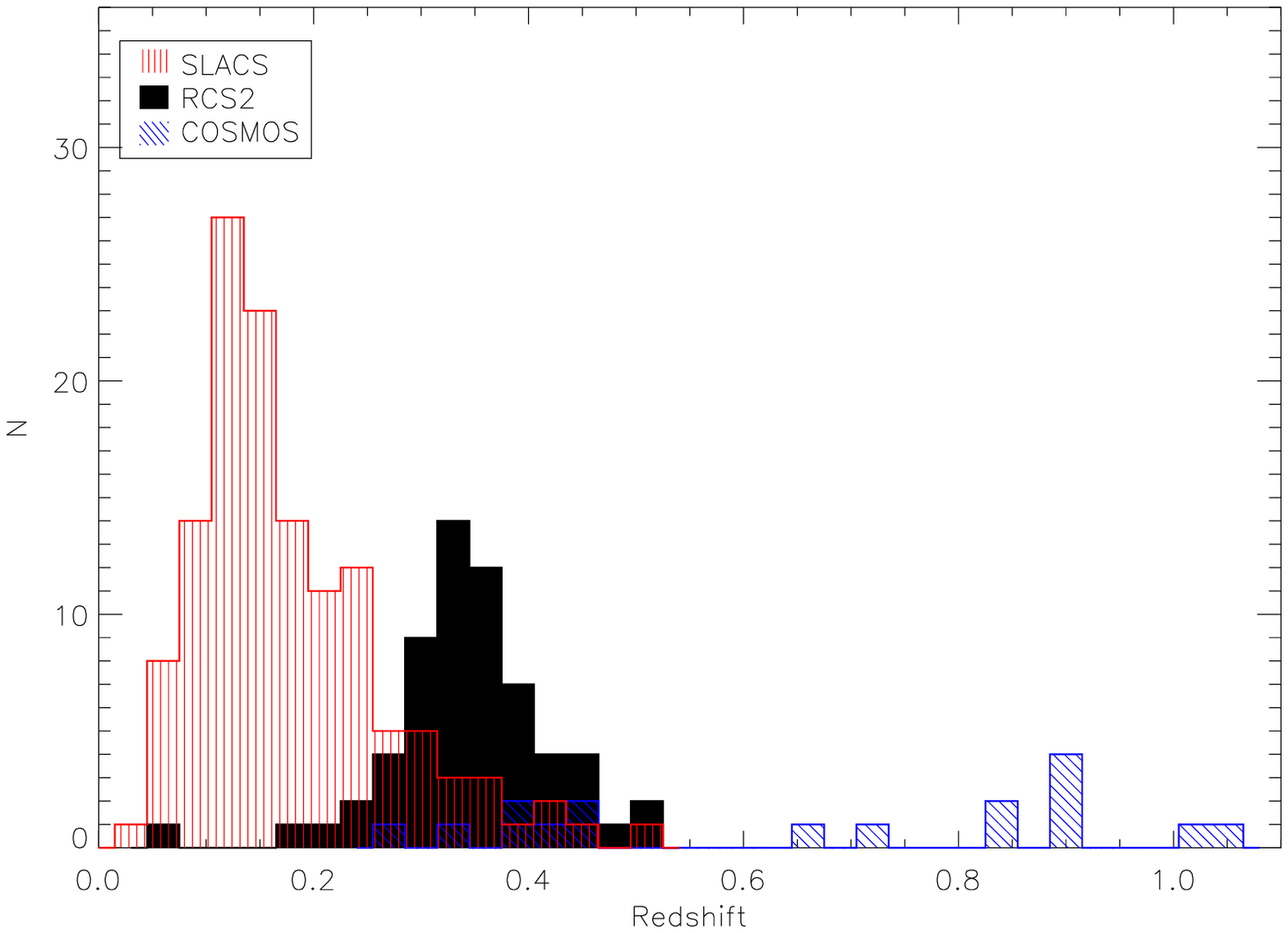}
 \caption{Redshift distribution of the RCS2 galaxy scale lenses compared to SLACS \citep{bolton08} and COSMOS lenses \citep{faure08}.}
 \label{fig:redrange}
\end{figure}

The properties of the RCS2 survey and the lens selection method have as a consequence some remarkable differences and advantages in comparison with other large galaxy galaxy lens samples:

\begin{itemize}
 \item \underline{\textit{Bright high-redshift lenses}}: All lenses in the catalog are either spectroscopically confirmed or photometric candidate Luminous Red Galaxies. This naturally results in bright lens galaxies that can reach high redshifts. In particular, the combination of the relatively shallow but wide photometric data from RCS2 and the selection of LRGs as deflectors, place the mean lens redshift of the sample in a currently under-sampled range. Figure \ref{fig:redrange} shows a comparison of the lens redshift distribution of different galaxy scale lens samples: SLACS \citep{bolton08}, on the lower redshift end with a median redshift of $z$=0.16, and COSMOS \citep{faure08}, on the higher redshift end with a median redshift of $z$=0.71. The RCS2 galaxy scale lens sample has a median redshift of $z$=0.34. Thus, this sample will allow us to fill the current gap of galaxy evolution studies through gravitational lensing.
 \item \underline{\textit{High-redshift sources}}: The higher redshift of the lens candidates coupled with the photometric selection of the arcs (high $g^\prime$-band brightness) does not constrain the redshift of the sources to low redshift, allowing source redshifts up to $z\sim$3.5 ($g^\prime$ band central wavelength=4872\AA{}). Harnessing the power of lensing as a natural telescope, this allows the study of star forming galaxies at redshifts $>$1.0. 
\item \underline{\textit{Bright arcs}}: As mentioned before, the imaging of the RCS2 survey, in particular the $g^\prime$-band imaging, has been carried out using short exposures on a 3.6 m telescope. Since the distorted arcs were selected visually by searching for bright features in this band, most arcs have high surface brightnesses, allowing easy ground and space-based follow-up observations.
\item \underline{\textit{No cluster galaxy bias}}: The lens candidates have been blindly selected over large areas based on their individual photometric and/or spectroscopic properties. Thus, to a first approximation there is no bias toward cluster galaxies in the sample in the selection process, making the sample homogeneous by construction (note however, that LRGs in general are likely to be found in cluster environments e.g., \citealt{deng09, sanchez09}).
\item \underline{\textit{Environment}}: Since the candidates were selected from the RCS2 fields, extensive photometric information on their surroundings is available and any clusters and/or large galaxy groups in their environments will be identified and characterized (paper in preparation). This, coupled with the ``no bias'' property, will allow us to contribute to the controversy of the relationship between lens and cluster galaxies (e.g., \citealt{faure09}).
\end{itemize}

\begin{figure}[h]
 \centering
 \includegraphics[clip=true, trim=30pt 0pt 0pt 0pt, width=8.75cm]{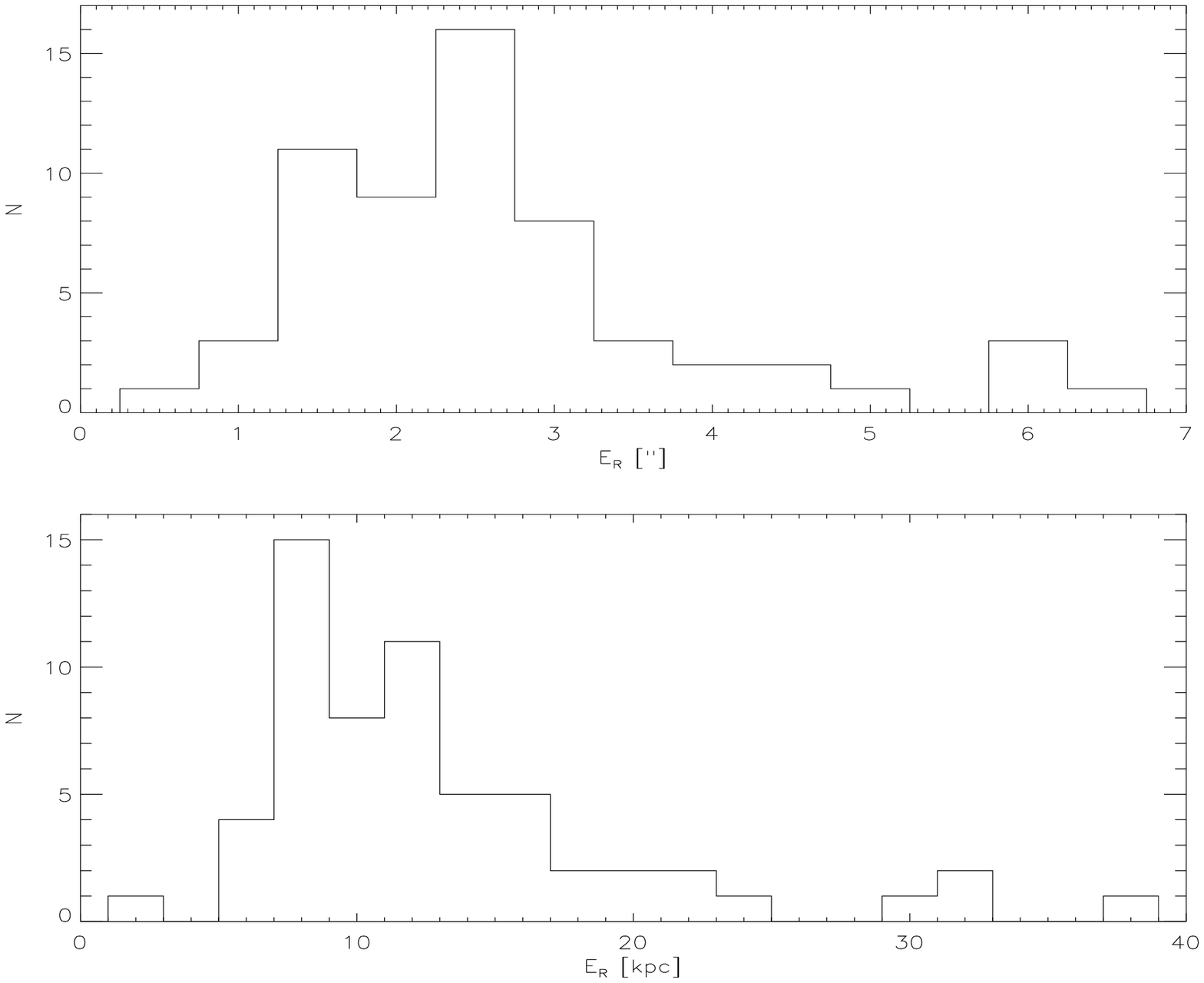}
 \caption{Einstein radii distribution of the RCS2 lens candidate systems in observed arcsec (top panel) and in kpc projected to the lens planes (bottom panel).}
 \label{fig:erdist}
\end{figure}

\section{Mass and Mass-to-light ratios}

The photometric data currently available do not allow the identification of distinct features in the strongly lensed arcs which would easily allow proper lens models of the systems to be constructed. However, a good approximation for the Einstein radii of gravitational lenses is given by half of the distance between opposite multiple images, or by the arc radii when no counter images are evident. The median measured Einstein radius of the systems in the catalog is of 2.5 arcsec. Projected to the lens planes, the median Einstein radius of the sample is of 11.5 kpc (see Figure \ref{fig:erdist} for the Einstein radii distributions).

Motivated by the results shown by \cite{gavazzi07}, who show that isothermal profiles are generally a sensible choice for galaxy lens mass distributions, we assume Singular Isothermal Sphere (SIS) profiles for the lenses in our catalog. The three dimensional mass enclosed within a certain radius assuming an SIS profile is obtained simply through (e.g., \citealt{SKW06})

\begin{equation}
M(R)=\frac{2 \sigma_v^2R}{G},
\end{equation}

\noindent where the SIS velocity dispersion,

\begin{equation}
\sigma_v^2=\frac{\theta_E c^2}{4 \pi} \frac{D_S}{D_{LS}},
\end{equation}

\noindent can be directly measured from the lensing geometry. $\theta_E$ is the Einstein radius (as measured from the arc radius or multiple images separation) and $D_L$, $D_S$ and $D_{LS}$ are the cosmological angular diameter distances between the observer and the lens, the observer and the source, and the lens and the source, respectively.

Obtaining accurate photometric redshift measurements for the sources in the catalog is not possible with the current data set. Even when the sources are bright, they are significantly fainter ($\sim$2 mag) than the very close lens galaxies in the sample precluding any sensible flux measurements. Additionally, and also due to the small distance between the lenses and the sources, photometric spectral energy distributions for the sources would be significantly biased due to the unknown extinction produced by the lensing galaxies. Thus, with the current data set we need to assume a most likely value for the sources' redshifts based on the underlying lensed galaxy distribution, independent of the individual photometric information. This value has been drawn from the models shown by \cite{bayliss11b}, which show the probability distributions for lensed sources in wide surveys. Selecting a 68\% confidence interval for a survey depth of $g$=25 (dashed-line distribution shown in Figure 2 of \citealt{bayliss11b}), we obtain an estimated 
redshift range of 0.8$<$$z_s$$<$2.0 with a most likely value of $z_s$=1.28. Using these values and the measured Einstein radii of the systems in the catalog, we estimated the Einstein radii enclosed three dimensional SIS masses for the lenses (using $R=\theta_E D_{L}$ in Equation (3)). These values range between 1.0$\times$10$^{\rm 10}$$M_{\odot}h^{-1}$$\lesssim$M$\lesssim$1.0$\times$10$^{\rm 13}$$M_{\odot}h^{-1}$, with a median enclosed mass of M=5.5$\times$10$^{\rm 11}$$M_{\odot}h^{-1}$.

\begin{figure}
 \centering
 \includegraphics[clip=true, trim=30pt 0pt 0pt 0pt, width=8.75cm]{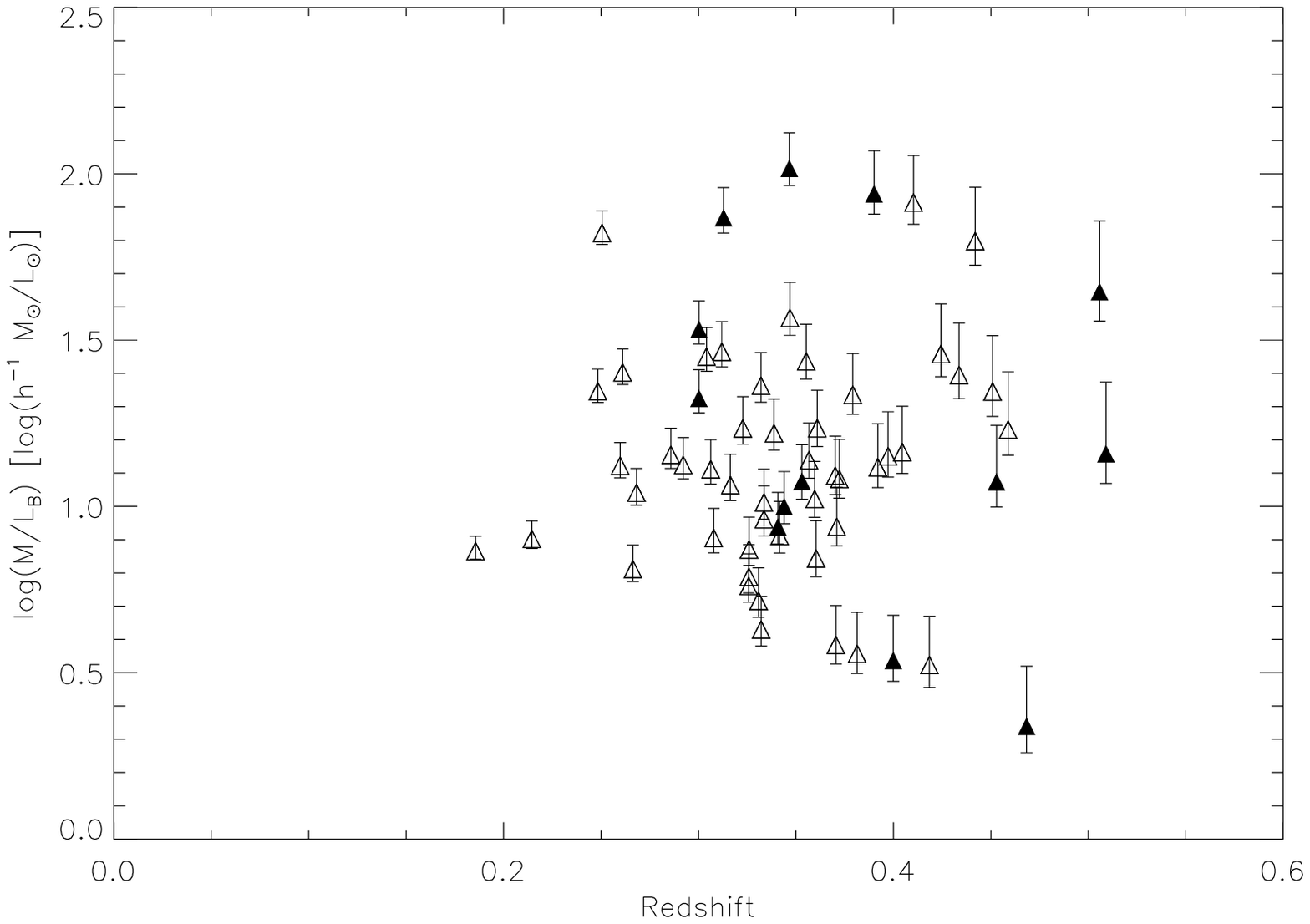}
 \caption{Mass-to-light ratios of the lenses in the catalog with spectroscopic (filled triangles) as well as photometric (open triangles) redshifts, assuming source redshifts of 1.28. The lower and upper error bars show the upper and lower redshift limits considered for the redshift of the sources (0.8$<$$z_s$$<$2.0), respectively.}
 \label{fig:ml}
\end{figure}

Coupling the enclosed masses with the RCS2 photometry of the lenses (and their redshifts [spectroscopic or photometric]) we obtain an estimate of the mass-to-light ratio of our sample, ranging from 1$M_\odot/L_{\odot,B}h\lesssim M/L \lesssim$120$M_\odot/L_{\odot,B}h$, with a median value of 14$M_\odot/L_{\odot,B}h$ (Figure \ref{fig:ml} and Tables \ref{tab:SDSSLRGs} and \ref{tab:RCSLRGs} show the mass and mass-to-light ratios for each lens).

\section{Summary and Discussion}

We have searched for galaxy scale lenses in the 820 deg$^{2}$ imaging from the RCS2 survey. We pre-selected LRGs as candidate lensing galaxies and searched for blue neighbors (arcs and/or multiple images) within 6 arcsec of the LRGs. The LRGs with neighbors were visually inspected taking advantage of the three-band imaging of the RCS2. The search yielded 60 candidate strong lensing systems with lens redshifts between 0.2$\lesssim z \lesssim$0.5, sampling a generally under-sampled redshift space for strong galaxy scale lensing. 

Due to the properties of the RCS2 survey and the search method the candidate lensing systems have some remarkable properties including high brightness of both lenses and sources, sources up to redshift $\sim$3.5 and no bias toward cluster galaxies. These properties will allow an accurate measurement of the evolution of the bright and dark mass profiles of LRGs as well as detailed morphological studies of the background sources via source reconstructions techniques.

We have estimated the lensing mass enclosed within the Einstein radii of our candidates as a means to estimate the evolution of the mass-to-light ratio of LRGs. Previous studies have only shown weak evolution of the dark matter fraction with redshift \citep{treukoop04,auger10,faure11,ruff11}, in agreement (within the errors) with our findings. However, the mass resolution attainable with the current information, does not permit significant constraints on the evolution, with no evolution also being consistent within the uncertainties. Significant improvements will be seen with follow-up spectroscopy and high resolution imaging which will confirm the nature of the candidate systems and allow robust lens mass modeling.

Searching for strong lenses by means of visual inspection is not a new technique (e.g., \citealt{fassnacht04,faure08,jackson08}), and has had so far higher success rate than fully automatic algorithms (e.g., \citealt{seidel07,alard06}). However, the method has two important drawbacks: first, it is limited to relatively shallow and/or narrow surveys where of the order of thousands of systems can be visually inspected and second, even though possible, it is hard to draw a quantifiable selection bias, hurting the lens number statistics and, thus, the use of lens counts to constrain cosmological parameters (e.g., \citealt{wambsganss08,fedeli08,hilbert09}). In the near future, with deep all-sky extragalactic surveys such as LSST, it is unrealistic to be able to create statically sound lens samples based on visual selection. It is therefore desirable that the lensing community moves toward fully algorithmic lens search methods to take full advantage of these huge data sets with detailed testing of incompleteness 
effects. Nevertheless, with current galaxy galaxy lens number counts, these (smaller volume) visually selected catalogs, are still relevant and an important source of new lens systems. Additionally, due to their high efficiency, visually selected samples serve as benchmarks and constraints to improve and calibrate automatic selection algorithms.

\acknowledgments

L.F.B. research is partially funded by Centro de Astrof\'isica FONDAP and by proyecto FONDECYT 1085286. This work is based on observations obtained with MegaPrime/MegaCam, a joint project of CFHT and CEA/DAPNIA, at the Canada-France-Hawaii Telescope (CFHT) which is operated by the National Research Council (NRC) of Canada, the Institute National des Sciences de l'Univers of the Centre National de la Recherche Scientifique of France, and the University of Hawaii.

\bibliographystyle{apj}
\bibliography{ANGUITA}

\clearpage

\end{document}